\documentclass{article}
\usepackage{graphicx} 
\usepackage{amsfonts}
\usepackage{amssymb}
\usepackage{amsmath}
\usepackage{ulem}
\usepackage{dirtytalk}
\usepackage{authblk}
\usepackage{hyperref}
\usepackage{appendix}
\usepackage[sort,compress,super,square]{natbib}
\newcommand{\gfrak}{\mathfrak{g}}

\newcommand*{\rbb}{\mathbb{R}}

\newcommand{\ccal}{\mathcal{C}}
\newcommand{\lcal}{\mathcal{L}}

\newcommand{\p}{\partial}
\newcommand{\s}{\Sigma}

\newcommand{\Tr}{\text{Tr}}

\title{Edge modes of tetrad gravity: Unlike diffeomorphisms, all shifts are integrable}
\author{Simon Langenscheidt }

\affil{LMU Munich, MQCST\footnote{S.Langenscheidt@lmu.de}
}
\date{\today}

\begin{document}

\maketitle

\begin{abstract}
    We present an improved notion of internal tetrad shifts in 4 dimensions which is always integrable in the presence of corners. This allows us to study the fully extended corner symmetry algebra of gauge charges, which is a deformation of $ISO(1,3)^S$ involving spacetime curvature. We argue this implies corner noncommutativity of the spin connection $\omega$. The latter in particular hints that an extended BF theory might be a better way understand the dynamics of tetrad gravity. This result presents us with an integrable, complete set of edge modes for gravity in 4D, with potential ramifications for asymptotic symmetries and quantisation.
\end{abstract}

\tableofcontents

\section{Introduction}
In a recent publication\cite{langenscheidtNewEdgeModes2025}, we studied an internal symmetry of the 4D tetrad formulation of GR, in particular Einstein-Cartan theory\footnote{Globally, we use geometrised units $8\pi G=1=c$ in this note. We also omit, apart from in appendices \ref{App:tech},\ref{App:Der},\ref{App:existence}, the distinction between spacetime connections $\omega$ and reduced connections $ \tilde{\omega}$. As we use a different reduction from the literature\cite{cattaneoReducedPhaseSpace2019}, we refer to appendix \ref{App:existence} for details on this point.}
\begin{equation}
    S[ e,\omega]= \int_M \epsilon_{IJKL} e^I\wedge e^J\wedge F_\omega^{KL}
\end{equation}
which, in analogy to their 3D gravity counterpart, are best referred to as \textit{shifts} and were mentioned before by a number of authors\cite{horowitzTopologyChangeGeneral1991,montesinosGaugeSymmetriesFirstorder2018,cattaneoReducedPhaseSpace2019}. These are parametrised by internal-space 4-vectors $\phi^I$ and shift the tetrad by a covariant derivative,
\begin{equation}
    Y_\phi[ e]^I = d_\omega \phi^I + \dots
\end{equation}
which makes the tetrad appear like a connection for the translation group $\rbb^{1,3}$. The main feature of these transformations is that they let us re-express diffeomorphisms as internal gauge transformations (meaning, they only depend on functions as parameters, rather than vector fields), but with transformation parameters that depend on the fields $ e^I,\omega^{IJ}$. Such a thing happens in many topological theories, for example Chern-Simons theory\cite{geillerEdgeModesCorner2017} or 3D gravity\cite{geillerDiffeomorphismsQuadraticCharges2022}. We pointed out then that the transformations we presented were not integrable in the presence of corners, meaning there was no generating function $P_\phi$ on the phase space of gravity that generated them in the form $Y_\phi[-] = \{P_\phi,-\} $. \\

Contrast this with the other common type of internal symmetries the theory has: Lorentz transformations which rotate different tetrads into each other,
\begin{equation}
    X_\alpha[e]^I=-(\alpha\cdot e)^I:=-\alpha^I{}_J e^J\quad X_\alpha[\omega]^{IJ}=d_\omega \alpha^{IJ}:= d\alpha^{IJ}+[\omega,\alpha]^{IJ}.
\end{equation}
For field-independent $\alpha$, this is integrable with generator\footnote{We use the notation $X_\beta^{IJ}= (\star+\beta)X^{IJ}= \beta X^{IJ}+\frac{1}{2}\epsilon^{IJKL} X_{KL}  $ including the internal Hodge star $\star$.}
\begin{equation}
    J_\alpha =\frac{1}{2}\int_\Sigma \frac{1}{2} (e^2_\beta)^{IJ} \wedge d_{{\omega}}\alpha_{IJ}
\end{equation}

The present note remedies this issue completely. We present that the original shifts, which we will refer to as the \textbf{naive shifts}, become fully integrable\footnote{For all field-independent parameters $\phi$.} if one augments them by a field-dependent Lorentz transformation. We will refer to these integrable shifts as the \textbf{improved shifts}.\\

These improved shifts come with advantages and disadvantages. The advantage is that the charges are well-defined regardless of boundary conditions, as well as their Poisson bracket. We can therefore present the algebra in full, and find that for its closure, we need to include \textbf{(a)} a corner charge given by a smearing of the curvature $\oint F_\omega$, \textbf{(b)} corner charges linear in the fields, $\oint  e,\oint\omega$. After its derivation, we will treat this as a fact, and discuss an interpretation of this result in terms of corner Poisson brackets of $\omega, e$. Furthermore, the presence of $F_\omega$ in the algebra can be derived from a certain (nonstandard) presentation of GR as a constrained topological field theory. We therefore learn many things about the corner and symmetry structure of tetrad gravity for free, in addition to having an integrable set of gauge transformations. \textit{I stress that this is not the case for generic diffeomorphisms which move the corner.}\\
The disadvantage is that the resulting transformations are highly non-explicit in the sense that one \textit{can} in principle give explicit expressions for them, they are just not particularly insightful. This will be particularly true for the correcting Lorentz transformation one needs to add to the naive shifts, whose parameter is always defined, but whose meaning is nebulous as of the time of this note. \\
A consequence of this is that the resulting calculations are quite intricate. What allows them to stay tractable is the integrability of the transformations, which guarantees that the algebra of \textit{charges} one can calculate is a good representation of the algebra of the \textit{transformations}. It also means that finding the finite form of the transformations is potentially very complicated, and at the time of writing it is unclear to me how to find them.\\

Nevertheless, these transformations are potentially very useful for characterising the corner data of tetrad gravity and many approaches related to discretisation, as well as asymptotic symmetries. Many of these applications were already discussed in related work\cite{langenscheidtNewEdgeModes2025}, but some more follow in the discussion. \\
The structure of this note is the following. We first present the result, and then delegate its verification, along with other calculations, to the appendix \ref{App:Var}. We then present the algebra of the corner charges and point out implications for the corner symplectic form of the theory (which is a notion we will introduce in place) in section \ref{Alg}. We then present an embedding of the 4D theory in an extended BF theory in section \ref{EmbIntoeBF} in which the algebra of the shifts look particularly natural. Finally we will point out some future directions in section \ref{Outlook}. Appendices will be pointed out in relevant places in the main text.

\newpage

\section{The result}\label{Result}
Our setting is the covariant-canonical off-shell phase space $\ccal_\s$ of ECH on a slice $\s$ (either spacelike or timelike)\footnote{Lightlike slices may be treated as well with additional technical complications\cite{canepaGravitationalConstraintsLightlike2021}.}, equipped with the with symplectic form\footnote{We use the convention for the matrix trace $\Tr[XY] = \frac{1}{2} X_{IJ}Y^{IJ} $. Indices are raised and lowered with the internal Minkowski inner product $\eta_{IJ}$.}
\begin{equation}
    \Omega = \frac{1}{2} \int_\Sigma \Tr[\delta e^2_\beta\wedge\delta {\omega}]= \int_\s \delta e_I\wedge\delta p^I,
\end{equation}
which allows a Darboux basis rewriting using the Brown-York momentum $p^I := -\omega_\beta^{IJ}\wedge e_J$.
We show here that the improved shifts
\begin{equation}
   \boxed{ Y_\phi[ e] = d_{ \omega}\phi - L[\phi]\cdot e \quad Y_\phi[{\omega}]\wedge e = d_\omega( L[\phi]\wedge e) - F_\omega\wedge\phi}
\end{equation}
are integrable for field-independent $\phi$, if we demand
\begin{equation}
     L[\phi]_\beta^{IJ}  e_J = p'(\omega^{IJ}_\beta \phi_J).
\end{equation}
$L:\ccal_\s\times \rbb^{1,3}\rightarrow \gfrak=\mathfrak{so}(1,3)$ is a field configuration-dependent mapping that turns a vector into a bivector, so an element of $\mathfrak{so}(1,3)$, and $p'$ is a certain projection operator. As we show in appendix \ref{App:existence}, it always exists with the stated property. The improved shifts are then generated by the charge
    \begin{equation}\label{ShiftCharge}
   \boxed{ P_\phi = - \int_\Sigma \Tr[ (\phi\wedge e)_\beta\wedge F_{{\omega}} + \frac{1}{2}d_{{\omega}} e^2_\beta\;  L[\phi] ] -\oint_{\partial\Sigma}p_I \phi^I }
\end{equation}

which is a typical gauge generator in the sense that its bulk part is a constraint, and therefore vanishes for physical configurations. 
The constraint in question is a combination of the Einstein constraint $(F_\omega)_\beta^{IJ}\wedge e_J $ and the Gauss constraint $\frac{1}{2}d_{{\omega}} e^2_\beta$, which are both zero on the physical phase space. On the boundary of the slice, the \say{corner}, there is however a piece that does not vanish. This is known as a \textit{corner charge}, and is presently
\begin{equation}
     e^2_\beta\cdot L[\phi] = - p_I \phi^I
\end{equation}
which is the Brown-York momentum, written in a covariant form.
We move the calculation verifying the result of integrability to the appendix \ref{App:Var}.
We can also rewrite the generator in a way where all terms are in the bulk, which then has a much more illuminating form:
\begin{equation}
    P_\phi = - \int_\s p_I \wedge d_\omega\phi^I + \phi_I (F_\omega-d_\omega\omega)^{IJ}_\beta \wedge e_J.
\end{equation}
In this form, we can see that it consists of a momentum-derivative pair $p_I d_\omega\phi^I$, just like the generator of Lorentz transformations does, but also an additional piece. This one involves the peculiar non-tensorial combination $F_\omega-d_\omega\omega = -\frac{1}{2}[\omega,\omega] $, which vanishes for \textit{reducible connections}, so precisely those where the commutator vanishes. It is then clear that the corner charge piece comes from the first term, and the bulk constraint is
\begin{equation}
    d_\omega p^I = (F_\omega-d_\omega\omega)^{IJ}_\beta \wedge e_J.
\end{equation}
This has the form of a (non)conservation law for the momentum $p$, with violations for non-reducible connections. \\
We can rewrite the transformation law of the connection in the following way:
\begin{equation}
    Y_\phi[{\omega}]_\beta^{IJ} \wedge e_J = -{\omega}_\beta^{IJ} \wedge d_\omega\phi_J.
\end{equation}
This shows that the momentum $p$ would be invariant under $Y_\phi$, if not for the Lorentz rotation in the transformation of $e$:
\begin{equation}
    Y_\phi[p]^I = {\omega}_\beta^{IJ}\wedge (L[\phi]\cdot e)_J 
\end{equation}
This is no coincidence: If, instead of the above terms with $\omega$, we had used some fixed background connection $\Bar{\omega}$ (i.e. $\delta \Bar{\omega} =0$), then the generator
\begin{equation}
    T_\phi = -\int_\s p_I\wedge d_{\Bar{\omega}}\phi^I
\end{equation}
would preserve $p$ and only change $e$ by a $\Bar{\omega}$-covariant derivative. However, this would \textit{not} be a gauge generator, as this is not the dynamics of gravity. The use of the Einstein constraint in the bulk in $P_\phi$ guarantees that $Y_\phi$, unlike the transformation generated by $T_\phi$, is a true gauge transformation of the theory. What this means is that the Brown-York momentum is \textit{not} an invariant quantity under the improved shifts.\\
The improved shifts, like the naive ones, have the feature of letting us rewrite diffeomorphisms as field-dependent gauge transformations:
\begin{equation}
    \lcal_\xi e^I = d_\omega i_\xi e^I + i_\xi (d_\omega e^I) - (i_\xi\omega\cdot e)^I
\end{equation}
which we derive via Cartan's magic formula, and similarly
\begin{equation}
\begin{aligned}
        (\lcal_\xi\omega)_\beta^{IJ}\wedge e_J = i_\xi( (F_\omega)_\beta^{IJ}\wedge e_J) - (F_\omega)_\beta^{IJ} i_\xi e_J 
        + d_\omega (i_\xi\omega)_\beta^{IJ}\wedge e_J 
\end{aligned}
\end{equation}
which derives from $\lcal_\xi \omega = i_\xi F_\omega+ d_\omega i_\xi \omega$. 
If we choose $\phi_\xi = i_\xi e$, $\alpha_\xi= i_\xi\omega - L[i_\xi e]  $, then we have the on-shell equivalence
\begin{equation}
    \boxed{\lcal_\xi \approx Y_{\phi_\xi} + X_{\alpha_\xi}}.
\end{equation}
This is a generic phenomenon in topological field theories, like Chern-Simons or BF\cite{geillerDiffeomorphismsQuadraticCharges2022}. It means in particular that we can reparametrize the gauge structure of 4D gravity in terms of Lorentz transformations and improved shifts. However: unlike diffeomorphisms, all shifts are integrable in tetrad gravity.

\section{Algebra of charges}\label{Alg}
Given the transformations, we can calculate the on-shell algebra of corner charges, i.e. evaluate Poisson brackets of generators on-shell:\footnote{Lie brackets of parameters refer to the $\mathfrak{so}(1,3)$ bracket.}
\begin{equation}
    \boxed{\begin{aligned}
        \{J_\alpha,J_\beta\} &= J_{-[\alpha,\beta]}\\
        \{J_\alpha,P_\phi\} &= P_{\alpha\cdot\phi} + \oint_{\partial\Sigma} \Tr[(\phi\wedge e)_\beta\wedge d\alpha]\\
        \{P_\phi,P_{\Tilde{\phi}}\} &\approx 
        J_{[ L[\phi],  L[\tilde{\phi}]]}+
        \oint_{\partial\Sigma} \Tr[ \omega\wedge d(\phi\wedge\Tilde{\phi})_\beta
        -(\phi\wedge\Tilde{\phi})_\beta\; F_\omega   ]
    \end{aligned}}
\end{equation}
The last equality (which is the only one where we explicitly invoke vanishing of the bulk constraints) in particular receives contributions from the bulk that must be carefully treated.\footnote{It involves expanding $\Tilde{\phi}_I d_\omega({\omega}_\beta\cdot\phi)^I-\phi_I d_\omega({\omega}_\beta\cdot\Tilde{\phi})^I+\Tr[(\phi\wedge\Tilde{\phi})_\beta\cdot F_\omega]$.} In addition to an expected Poincare-structure, we get extension terms which are in principle a little complicated due to field-dependence, meaning they are not central. The off-shell result for the shifts is in equation \ref{OffshellBracket}, and a derivation of this algebra is in appendix \ref{App:Commutator}. \\
Note that we can rewrite this new term as
\begin{equation}
    \oint_{\partial\Sigma} \Tr[  \omega\wedge d(\phi\wedge\Tilde{\phi})_\beta
        -(\phi\wedge\Tilde{\phi})_\beta\; F_\omega ] = -\oint_{\partial\Sigma} \Tr[ \frac{1}{2} [\omega,\omega]_\beta\;(\phi\wedge\Tilde{\phi})]
\end{equation}
so the term vanishes only for reducible connections.\\
Taken purely on-shell, these brackets imply certain commutation relations for the fields $e,\omega$ on the corner. In particular, they imply that tetrads do not commute on corners, a result that has been well-known for a while\cite{freidelLoopGravityString2017,freidelBubbleNetworksFramed2019,freidelEdgeModesGravity2020,freidelEdgeModesGravity2021}. What is new here is that we can also deduce that the same holds for the connection. In particular, we can deduce from the shift-shift Poisson bracket that the objects $Q_\phi^I := \omega_\beta^{IJ}\phi_J$ should satisfy
\begin{equation}\label{ConnNC}
    \{Q^I_{\phi,a}(x),Q^J_{\psi,b}(y)\} = \frac{1}{2}\epsilon_{ab}\delta(x,y) [ L[\phi], L[\psi]]^{IJ}_\beta(x).
\end{equation}
This is due to expanding out the Poisson bracket of two $p^I$'s and matching it with the right hand side of the shift-shift bracket in terms of powers of $e$ and $\omega$. Similarly, the bracket of $\omega,e$ is highly constrained, but likely nonzero. We do not yet claim to have the precise Poisson structure at hand yet; we leave this to future work. \\
What is most important in this argument is the qualitative conclusion: The general corner charge structure of tetrad gravity requires the corner connection to be in the phase space, and also to have certain \textit{nontrivial commutation relations with itself and the tetrad}.\\
In particular, if we take the rough shape of equation \ref{ConnNC} as the expected result, then we have a sort of Chern-Simons like noncommutative connection 
\begin{equation}
    \{A_a^i, A_b^j\} = \epsilon_{ab} \mathcal{M}^{ij}
\end{equation}
where for the gravity case $\mathcal{M}$ is field-dependent. This is very interesting as it suggests that a \say{deformed} Chern-Simons like phase space may be naturally present in 4D tetrad gravity. This is supported by the corner symplectic structure known for the tetrad\cite{freidelEdgeModesGravity2021}, and fits with studies of gravity in the context of quasilocal horizons\cite{engleBlackHoleEntropy2010,Bodendorfer:2013jba} in tetrad gravity.

\section{Embedding into extended BF theory}\label{EmbIntoeBF}

The fact that the shifts are noncommuting may surprise many readers expecting a proper Poincaré or (Anti)deSitter algebra. The appearance of the curvature term is, luckily, geometrically natural as curvature measures the noncommutativity of translations in a spacetime. Therefore, Poisson noncommutativity of shifts, if anything, \textit{should} be measured by curvature.\\
However, we will provide another partial argument for the noncommutativity, which simultaneously gives an idea for settings where the shifts are easier to handle. The idea is to use the common logic of embedding tetrad gravity into a topological BF-like theory using additional restrictions known as \textit{simplicity constraints}. There is, in fact, a useful \textit{extended} BF theory whose symmetry structure is quite reminiscent of the one presented here, and which is equivalent to gravity with a certain bulk potential.\\
The extended BF theory is defined by the Lagrangian
\begin{equation}
    L_0 = t_I\wedge d_\omega e^I +  \Tr[ B\wedge F_\omega + \lambda\wedge d_\omega\pi -\lambda \wedge ( B-\frac{1}{2} e^2_\beta )]
\end{equation}
and is explored in more detail in appendix \ref{App:eBF}. Here, we only need to know the following: the theory is on-shell equivalent to tetrad gravity, and its phase space reduces to that of gravity when we impose $\lambda=F_\omega, t=0$. It also has six gauge generators: Two of these, $T_\phi,R_\beta$, satisfy
\begin{equation}
    \boxed{
    \{T_\phi, T_{\tilde{\phi}}\}= R_{(\phi\wedge\tilde{\phi})_\beta}.
    }
\end{equation}
 What now helps us understand the shift algebra is that the role of the shifts is similar to that of the $T_\phi$, and their Poisson bracket gives $R$, which, onshell of $F=\lambda$, reads
\begin{equation}
    R_\beta= \oint_{\p\s}\Tr[F_\omega \, \beta].
\end{equation}
This is clearly the type of curvature term we get from the shift algebra!\\
This suggests that we can embed the tetrad gravity shifts into this extended BF theory. Indeed, with the use of two of the remaining three generators, $K_\mu,\tau_v$ (see appendix \ref{App:eBF} for details), we can define a differentiable function\footnote{
$K^{(b)}$ denotes only the bulk piece of the curvature constraint, $K_\mu$.}
\begin{equation}
\begin{aligned}
        P_\phi &= T_\phi - \tau_{\omega_\beta\cdot\phi} + K^{(b)}_{(\phi\wedge e)_\beta} \\
        &= -\int_\s t_I\wedge d_\omega\phi^I + e_I\wedge d_\omega(\omega_\beta\cdot\phi)^I + \Tr[(\phi\wedge e)_\beta\wedge F_\omega]
\end{aligned}
\end{equation}
which reduces, on the tetrad gravity phase space, to the improved shift generator. \\
On the extended BF phase space, this generates the flow
\begin{equation}
\begin{gathered}
        Y_\phi = d_\omega\phi \frac{\delta}{\delta e} 
        + ((F_\omega)_\beta \phi-d_\omega(\omega_\beta\cdot\phi)) \frac{\delta}{\delta t}\\
        +( (d_\omega\phi\wedge e)_\beta + t\wedge\phi +[( e\wedge\phi)_\beta , \omega])\frac{\delta}{\delta B} 
\end{gathered}
\end{equation}
which shifts $ e,t$ in very familiar ways\footnote{\begin{equation}\begin{aligned}   
\delta P = \int_\s& 
    -\delta t_I\wedge d_\omega\phi^I
    + \delta e_I\wedge ((F_\omega)_\beta\cdot \phi-d_\omega(\omega_\beta\cdot\phi))^I\\
    &- \Tr[( (d_\omega\phi\wedge e)_\beta + t\wedge\phi +[( e\wedge\phi)_\beta , \omega]) \wedge \delta\omega]\end{aligned}\end{equation}}.
From this, we can see that the Poisson brackets take the form
\begin{equation}
    \{P_\phi, P_{\tilde{\phi}} \} = 
    -\int_\s 
    \Tr[F_\omega \wedge d_\omega(\tilde{\phi}\wedge\phi )_\beta]
    +d_\omega(\phi\cdot\omega_\beta)_I\wedge d_\omega\tilde{\phi}^I +  d_\omega\phi_I\wedge d_\omega(\omega_\beta\cdot\tilde{\phi})^I
\end{equation}
and can be calculated with significantly less effort compared to the ECH setting. The result is symbolically similar to the Poisson brackets of $T,T$ and $T,\tau$(see appendix \ref{App:eBF}). We also see certain terms are analogous to the terms we see in the bulk Poisson bracket. Most importantly, though, we see that this Poisson bracket has a natural corner piece
\begin{equation}
    -\oint_{\p\s} 
   \Tr[ F_\omega \,(\tilde{\phi}\wedge\phi )_\beta + \omega_\beta\wedge d( \phi\wedge\tilde{\phi} ) ]
\end{equation}
and thus explains on a much simpler level from where the noncommutativity arises: the $F$-term is due to the $T-T$ Poisson bracket producing $R$, and the $\omega$-term due to the $T-\tau$ bracket. \\
We can then also check that the combination $P$ is actually invariant under the most important of the extended BF constraints, the modified simplicity constraint $MSC_\Delta$:
\begin{equation}
    \{MSC_\Delta, P_\phi\}
    =
    -\int_\s t_I \wedge \Delta^{IJ}\phi_J - \Tr[(e\wedge \omega_\beta\cdot\phi) \wedge \Delta] 
\end{equation}
While this is not invariant under \textit{all} flows of the simplicity constraints, on the $t=0$ surface, we have that it \textit{is} invariant under flows with $\Delta_\beta^{IJ}\wedge e_J = 0 $, which are precisely the $\Delta$ that preserve the condition $t=0$. 
To see this, note that
\begin{equation}
	e\wedge (\omega_\beta\cdot\phi)= e\wedge p'(\omega_\beta\cdot\phi)=e\wedge (L[\phi]_\beta\cdot e) = [\frac{1}{2}e^2_\beta,L[\phi]].
\end{equation}
Then, identities from appendix \ref{App:Id} give that
\begin{equation}
\begin{aligned}
		\Tr[(e\wedge \omega_\beta\cdot\phi)\wedge \Delta] &= \Tr[[\frac{1}{2}e^2_\beta,L[\phi]]\wedge \Delta] = \Tr[[\Delta_\beta,\frac{1}{2}e^2]L[\phi] ] \\
		&=
	\Tr[  (\Delta_\beta\cdot e\wedge e)  L[\phi] ]  = 0
\end{aligned}
\end{equation}
So, the function $P_\phi$ is at least invariant under simplicity constraints on the $t=0$ surface, and can therefore be understood as a residual gauge transformation of extended BF theory on the phase space reduced by $K,MSC$ and $t=0$\footnote{More generally, they are preserved by all field-dependent $\Delta$ such that $t_J\wedge \Delta^{JI} + e^K \wedge \Delta_{KJ}\wedge \omega_\beta^{JI}=0  $. }. This is precisely the ECH pre-phase space. \footnote{To see that they preserve the simplicity constraints in the opposite calculation of the Poisson bracket, one needs to be careful: One needs to apply $Y_\phi$ to specifically $MSC_{\Bar{\Delta} }$ with $\Bar{\Delta}_\beta^{IJ}e_J=0$. Evaluating the Poisson bracket off-shell, and going to the ECH pre-phase space, it once again vanishes after using some identities. } \\
Therefore, it stands to reason that the extension of BF theory we presented gives a context in which the given shift symmetries can be studied with simpler and more explicit calculations, and in which their meaning is more obvious.
To summarise this argument: While the improved shifts on the tetrad phase space are seemingly complicated, there are more natural pendents of them on an extended phase space. In this extended BF phase space, we can also understand the origin of their noncommutativity, and corner charges linear in the fields $e,\omega$ appear naturally. 

\section{Outlook}\label{Outlook}

The results reported in this note are quite unexpected. Given the complexity of diffeomorphisms, and the generic non-integrability of the naive shifts, is it quite remarkable that there are integrable, full sets of gauge transformations like that given by Lorentz transformations and improved shifts. This informs us that there is something quite special about the structures we found, even if the details of interpretation are still a little unclear at this point in time.\\
We already reported on several possible applications regarding the internal shift symmetries in the prequel\cite{langenscheidtNewEdgeModes2025}. In this outlook, we will instead consider the additional questions and insights that come up due to the improved shifts, particularly their integrability and complexity.\\

First, it should be noted that while we did not consider a cosmological constant or matter in the dynamics here, this is not a matter of principle. The expressions of the bulk shift constraint may change, but the only problems arise from bulk derivative terms, which can at worst come from matter contributions. These have to be studied carefully, and are generically complicated. A cosmological constant, however, is no issue by itself.\\
Another obvious point is the interpretation of the Lorentz element $L[\phi]$. While we know it exists and is an ultralocal function of the fields, it begs for a geometric interpretation which we are lacking at the moment. One particular direction of attack on this matter may be to interpret it as part of the larger symmetry structure of the theory: it is perhaps to be understood as a mapping
\begin{equation}
    \rbb^{1,3}\rightarrow \text{Aut}(\rbb^{1,3})
\end{equation}
which is needed in the definition of a crossed module (also known as a 2-group). An analogous structure is present in 3D gravity\cite{Dupuis:2020ndx} as $v^i \mapsto \epsilon^{ijk}v_k $ and leads to quantum group structures at the quantum level. In 4D, we seem to have a symmetry structure 
\begin{equation}
    \mathfrak{so}(1,3)\ltimes (\rbb^{1,3}\times_c \mathfrak{so}(1,3)^\ast)
\end{equation}
which is a more general structure than the bialgebras $\mathfrak{h}\ltimes \mathfrak{h}^\ast $ that appear in 3D. Therefore, the situation may end up quite different than in lower dimensions.\\
A natural extension of this question is to better understand the role of the extended BF theory and its shifts. This theory may be more amenable for the construction of certain quantisations, like in state sum models\cite{perezSpinFoamApproach2013}. It is particularly of interest whether or not there is a similar way to get $P_\phi$ from extended BF theory in the discrete. This would circumvent a lot of complicated phase space problems and enable a more straightforward quantisation.\\

However, this is all but a prelude to the really interesting questions. \\
Since the shifts are generically integrable, they have a much better status than diffeomorphisms. Therefore, their application in the study of asymptotic symmetries is bound to bring to light many new insights about the symmetry structure of the gravitational S-matrix. In particular, most of the ambiguities that plague the study of diffeomorphism charges at infinity\cite{Speranza:2022lxr,barnichSurfaceChargeAlgebra2008,Freidel:2021cjp} are \textit{avoided ab initio} by the full integrability we found. It will be very exciting to study the symmetries and their relation to existing, ever larger symmetry algebras like Virasoro and the Weyl-BMS algebra\cite{freidelWeylBMSGroup2021}.\\
Of course, the implications of the symmetry structure go beyond the S-matrix: \textit{Any} region of gravity, be it finite or asymptotic, will be equipped with the symmetry algebra \ref{Alg}. This means that the study of this algebra and its resulting representation theory brings with it crucial information about the Hilbert space of quantum GR on any bounded region. This is particularly interesting for building up such Hilbert spaces from cellular decompositions, in the spirit of topological field theories\cite{cattaneoCellularTopologicalField2020} or spin networks\cite{baezSpinNetworksNonperturbative1998, freidelCornerSymmetryQuantum2023}, where these corner charge algebras govern the glueing of degrees of freedom between cells.\\
A particularly immediate impact on these approaches is the presence of linear charges $e,\omega$ on the corner. The continuum symmetry structure hints here at the necessity to include such charges or variables, which would amount to the presence of discretised tetrad and connections (i.e. edge vectors and Wilson lines) on the \textit{boundaries of or intersections between} 3D cells. The precise structure of these corner degrees of freedom needs to be studied in much more detail. What can already be surmised at this level is that there are certain analogies to the phase space structure of Chern-Simons theories, but these need to spelt out in much higher precision.

\section*{Acknowledgements}
The author would like to thank Laurent Freidel for giving an initial suggestion that led to this result. He also thanks Giulio Neri and Fabio Mele for helping confirm the integrability calculation and a number of identities.
The author would also like to thank the Perimeter institute for hospitality. Research at the Perimeter Institute is supported in part by the Government of Canada through NSERC and by the Province of Ontario through MEDT.

\appendix

\section{Useful Identities}\label{App:Id}
We present here some necessary identities for (bi)vector (= antisymmetric matrix) valued differential forms.
First, note that matrix transposes work as usual, but with form grading: The matrix product
\begin{equation}
	(A\wedge B)^{IJ}:= A^{IM}\eta_{MN}\wedge B^{NJ}
\end{equation}
satisfies
\begin{equation}
	(A\wedge B)^t = (-1)^{ |A|\, |B| } (B\wedge A)
\end{equation}
which lets us write the commutator
\begin{equation}
	[A,B]^{IJ}:= A^{IM}\eta_{MN}\wedge B^{NJ}-(-1)^{ |A|\, |B| } B^{IM}\eta_{MN}\wedge A^{NJ}
\end{equation}
as 
\begin{equation}
	[A,B]= (A\wedge B) - (A\wedge B)^t.
\end{equation}
Another crucial identity\cite{cattaneoReducedPhaseSpace2019} (see lemmata 3.4, 3.5 of the reference) we use in many places is 
\begin{equation}
	\star[A,B]=[\star A,B]=[A,\star B]
\end{equation}
for all antisymmetric $A,B$ of all form degrees. 
This implies the nontrivial
\begin{equation}
	\star[A,B] = 2\star(A\wedge B) = (\star A\wedge B)-(\star A\wedge B)^t
	= ( A\wedge \star B)-( A\wedge \star B)^t
\end{equation}
and in particular for $A=B$ being $1$-forms:
\begin{equation}
	\star[A,A] = 2\star(A\wedge A) = (\star A\wedge A)+( A\wedge \star A).
\end{equation}
For a general bivector $A^{IJ}$ and vectors $X^I$, $Y^J$, we have the commutator expansion
\begin{equation}
	[A,X\wedge Y] = (A\cdot X)\wedge Y + (-1)^{|A|\, |X|} X\wedge (A\cdot Y)
\end{equation}
so the commutator with $A$ acts as a derivation of degree $|A|$ over the wedge product.\\
Our convention for the trace is 
\begin{equation}
	\Tr[A\wedge B] = \frac{1}{2} A^{IJ}\wedge B_{IJ}= -\frac{1}{2} A^{I}_J \wedge B^{J}_I = -\frac{1}{2}(A\wedge B)^I_I
\end{equation}
Then, we have the following identity for all bivector-valued forms $C$:
\begin{equation}
	\text{Tr}[C A B] = \text{Tr}[C \frac{1}{2}[A ,B]] 
\end{equation}
which we derive as follows:
\begin{equation}
	\begin{aligned}
		\Tr[C \wedge A\wedge B] &= - \Tr[\star C \wedge \star(A\wedge B)] \\
		&= -\Tr[\star C \wedge \frac{1}{2}\star[A,B]]
		=\Tr[ C \wedge \frac{1}{2}[A,B]]
	\end{aligned}
\end{equation}

This allows us to prove (recall $A_\beta := (\star+\beta)A$)
\begin{equation}
	\Tr[C\wedge A\wedge B_\beta]=\Tr[C\wedge A_\beta \wedge B]=\Tr[C_\beta \wedge A \wedge B]
\end{equation}
by applying the above identities successively. A special case of this is for $C= a\wedge \phi$, with $\phi$ a 0-form vector, from which follows the identity
\begin{equation}\label{SwapBetaTriple}
	a_I \wedge (A\wedge B_\beta)^{[IJ]}\phi_J
	=
	a_I \wedge (A_\beta\wedge B)^{[IJ]}\phi_J.
\end{equation}

\section{Phase space technicalities}\label{App:tech}
As already expanded on in previous work\cite{langenscheidtNewEdgeModes2025}, we need to remember that on the phase space of a slice $\s$, connections always have to be reduced by precisely 6 degrees of freedom compared to their covariant spacetime form\cite{cattaneoReducedPhaseSpace2019}. The reduction is essentially just that one needs to impose some parts of the torsion constraint even to just set up the phase space.\\
The simplest, non-technical way one can see that this is necessary is by recognising that the conjugate variable to $e^A_i$ must be some function of $\omega^{AB}_i$ and $e$. Since $e$, on a slice, has $3\times 4 = 12$ components, this means that $\omega$ by itself, which has $3\times 6=18$ components, carries information that will not go into the symplectic partner of $e$. Therefore, one needs to \textit{reduce the connections} on the phase space by $6$ constraints to accomodate for this difference.\\

The redundancy in question, in the local coordinates we chose, is essentially the following: in the spatial components of the connection, one can dualise to get a matrix 
\begin{equation}
    B_{ci} := \frac{1}{2}\epsilon_{abc} \omega^{ab}_i.
\end{equation}
The redundancy in the symplectic form $\Omega$ of ECH that needs to be fixed is given by the gauge invariance
\begin{equation}
    B_{ci}\mapsto B_{ci} + \Bar{\Delta}_{(ci)}
\end{equation}
which shifts this component by a symmetric $3\times 3$ matrix. More generally, it corresponds to shifts $\omega\mapsto\omega + \Bar{\Delta}$ where $\star\bar{\Delta}^{IJ}\wedge e_J =0$ constrains the gauge parameter 1-form. This has as a consequence that \textit{the symmetric part of $B_{ci}$ is not observable or even has a phase space variable}. In particular, certain parts of curvature and torsion are not part of the physical data of gravity on a slice. \\
We know of two ways to fix this gauge: The first was constructed by Schiavina and Cattaneo\cite{cattaneoReducedPhaseSpace2019,canepaGravitationalConstraintsLightlike2021} and amounts to the following: If $\gamma[e]$ is the spacetime Levi-Civita connection, then their \textit{structural constraint} says that reduced connections $\tilde{\omega}$ satisfy
\begin{equation}
    B_{(ci)}(\tilde{\omega})= B_{(ci)}(\gamma[e]),
\end{equation}
so it sets the symmetric part of $B$ \textit{to that} of the Levi-Civita connection. This is chosen precisely so that there is a 1-to-1 correspondence between solutions of the torsion constraint $d_{\tilde{\omega}}e=0  $ and the Gauss constraint $ d_{\tilde{\omega}}e^2=0 $ (which happens to be invariant under the redundancy) on a slice.\\
The formal statement of the constraint involves the choice of a once-and-for-all-chosen kinematical, field-independent internal normal $\nu^I$ that completes the image of $e|_\s$ to a 4D basis. $\nu$ is a parameter for the gauge fixing, but its specific choice is per se irrelevant. It being fixed, the structural constraint says that a specific component of the torsion must take a certain form:
\begin{equation}
	\nu \wedge d_{\tilde{\omega}}e = \tau\wedge e
\end{equation}
Here, $\tau^I$ is any vector valued 1-form. Requiring that this has a solution $\tau(e,\tilde{\omega})$ fixes a 6D cokernel to be trivial, and this precisely fixes the redundancy. Let $p'_{1,2}$ be the projection to the complement of the kernel of the map $X\mapsto X\wedge e$. Then, the structural constraint says equivalently
\begin{equation}
	p'(d_{\tilde{\omega}}e)= d_{\tilde{\omega}}e.
\end{equation}
So the structural constraint ensures that the Gauss constraint function $d_{\tilde{\omega}}e\wedge e$ is always nonzero, and this means all components of the Gauss constraint are nontrivial constraints on $\tilde{\omega}$.\\
So the restriction on connections is really one on components of the torsion. Consider the spatial components $T^a$ of the torsion; generically, it admits an expansion\footnote{This may be seen most easily by using an adapted time normal $u_\s$, which is orthogonal to the slice triad\cite{langenscheidtNewEdgeModes2025}.}
\begin{equation}
    T^a = d e^a + R\wedge e^a + S^a_b E^b \qquad E^a = \frac{1}{2}\epsilon^{abc}e_b e_c 
\end{equation}
where $R$ is a 1-form representing vectorial torsion (3 degrees of freedom) and $S_{ab}$ a symmetric $3 \times 3$ matrix encoding both axial and traceless torsion (1 and 5 degrees of freedom, respectively). For the Levi-Civita connection $\gamma[e]$, $R(\gamma[e]),S^a_b(\gamma[e])$ are such that the right hand side vanishes identically, $T(\gamma[e])=0$, and are determined as functions of $e$. Then, the structurally constrained representative, instead, has a fixed $S^a_b$-piece, so that
\begin{equation}
    T(\tilde{\omega})^a = d e^a + R\wedge e^a +S^a_b(\gamma[e])\, E^b=  (R-R(\gamma[e]))\wedge e^a .
\end{equation}
This means that on the structurally constrained kinematical phase space $\ccal_\s$, parts of the torsion constraint are already imposed. These restrict only the spatial torsion, and fix only axial and traceless parts, leaving the vectorial torsion completely free, still.\\
For reference, if we were to include Dirac spinors in the theory, there would be a source purely for \textit{axial} 4D torsion, through a the axial spin current $J^I_{(A)} =\Bar{\psi}\gamma^I \gamma^5 \psi$\cite{mercuriFermionsAshtekarBarberoConnection2006}. Then, to work on the ECH phase space, one would need to amend the structural constraint to include the spatial part of this axial current, i.e.
\begin{equation}
    B_{(ci)}(\tilde{\omega})= B_{(ci)}(\gamma[e] + \kappa[J_{(A)}] ).
\end{equation}
To summarise, in order to turn the pre-phase space of $\omega,e$ into a phase space, one must impose parts of the Torsion constraint from the very beginning.

\section{Derivation of $P_\phi$}\label{App:Der}
Here, we give a simple way to arrive at the expression \ref{ShiftCharge}. The starting point is the weighted Einstein constraint
\begin{equation}
    -\phi_I (F_\omega)_\beta^{IJ}\wedge e_J 
\end{equation}
which, together with the Gauss constraint $\frac{1}{2} d_\omega e^2_\beta $, or its vacuum equivalent, the torsion constraint $d_\omega e$, reduce the space of $(\omega,e)$s to the proper physical boundary of gravity. Asking for a differentiable version of this constraint leads to $P_\phi$. As it stands, the above is \textit{not} differentiable: the variation of the connection gives a boundary term
\begin{equation}
    -\phi_I (\delta F_\omega)_\beta^{IJ}\wedge e_J = -\phi_I (d_\omega\delta\omega)_\beta^{IJ}\wedge e_J = -d( \phi_I\delta\omega_\beta^{IJ}\wedge e_J) + \dots
\end{equation}
which can not be matched with any term in the contraction of the symplectic form, $-I_{Y_\phi} \Omega_\s$. This is what it would mean to be functionally differentiable. For this purpose, we add the minimal term that cancels the boundary variation, together with a generic derivative-free term $s_\phi$:
\begin{equation}
   - \phi_I (F_\omega)_\beta^{IJ}\wedge e_J
   - e_I \wedge d_\omega( \omega_\beta^{IJ}\phi_J) + s_\phi
\end{equation}
This solves the differentiability problem; we must now choose $s_\phi$ so that the bulk piece is again a linear combination of constraints. To see what this is, we partially integrate the second term, giving a boundary piece $d(e_I\wedge \omega_\beta^{IJ}\phi_J)$ and bulk
\begin{equation}
	- \phi_I (F_\omega)_\beta^{IJ}\wedge e_J - d_\omega e_I \wedge  \omega_\beta^{IJ}\phi_J + s_\phi
\end{equation}
Now we only need to be careful with the torsion $d_\omega e$: If we treat the manipulations so far as being on the level of spacetime forms, then obviously $d_\omega e=0$ holds on shell in absence of spin sources. Then, there is nothing else we need to add in $s_\phi$. However, if we want to construct a generator on a slice $\s$, then the torsion constraint \textit{splits} into the Gauss constraint and structural constraint, as described in appendix \ref{App:tech}. Then we must replace the connection $\omega$ by the reduced $\tilde{\omega}$ in the above, and note the following, with language and notation adopted from Cattaneo and Schiavina\cite{cattaneoReducedPhaseSpace2019}: In
\begin{equation}
	- d_{\tilde{\omega}}e_I \wedge  \tilde{\omega}_\beta^{IJ}\phi_J, 
	\end{equation} 
one can identify $\omega_\beta^{IJ}\phi_J$, through the wedge and inner product $\eta_{IJ}$, as an object in the dual space of the space that $d_{\tilde{\omega}}e $ belongs to.  This means that fact, we can project using $p'$ to 
\begin{equation}
	- p'(d_{\tilde{\omega}}e)_I \wedge  p'(\tilde{\omega}_\beta^{IJ}\phi_J)
	=
	- d_{\tilde{\omega}}e_I \wedge  p'(\tilde{\omega}_\beta^{IJ}\phi_J).
\end{equation}
This is crucial, as it ensures the existence of the Lie Algebra element $L[\phi]$, see \ref{App:existence}. With these caveats, also here we can see $d_{\tilde{\omega}} e=0$ as a proper constraint, and in fact rewrite this term as
\begin{equation}
	- d_{\tilde{\omega}}e_I \wedge  L[\phi]_\beta^{IJ} e_J = -\Tr[ \frac{1}{2}d_{\tilde{\omega}}e^2_\beta \, L[\phi] ].
\end{equation}
This makes it clear that this differentiable extension of the Einstein constraint really is an extension by the Gauss constraint.

\section{Existence of $L(\phi)$}\label{App:existence}
Here, we establish the existence of the Lie algebra element $L[\phi]\in \mathfrak{so}(1,3)^\s$ defined by the relation
\begin{equation}
    L[\phi]\wedge e = p'(\omega\wedge\phi) \Leftrightarrow L[\phi]_\beta^{IJ} e_J = p'(\omega_\beta^{IJ}\phi_J) .
\end{equation}
In it, for the moment, we treat $\omega$ as a generic bivector-valued 1-form, and include a projection $p'$, points to which we need to return in a moment.
This is an equality on a fixed slice $\s$ and can be solved by pure linear algebra arguments. They are not particularly illuminating, but provide explicit expressions for $L$. We do not pay close attention to the numerical prefactors here, as our only point is to establish existence.\\
Since the equation is ultralocal, we can solve it pointwise. The equation is between trivector-valued $1$-forms, and we solve it in local coordinates $x^i$ and in a local internal frame. Expanded out, the equation reads ($L$ instead of $L[\phi]$ for readability)
\begin{equation}\label{EqForL}
    L^{[AB} e_i^{C]} = X_i^{ABC}:= p'(\omega \wedge\phi)^{ABC}.
\end{equation}
We now distinguish between two cases. First, consider when $[ABC]$ does not contain the time direction $0$. then, we have
\begin{equation}
    L^{[ab} e_i^{c]} = X_i^{abc}\Leftrightarrow \epsilon_{abc} L^{ab} e_i^{c} = \epsilon_{abc} X_i^{abc}
\end{equation}
so we have a single equation by dualising with the 3D Levi-Civita symbol. Introducing the inverse triad $\hat{e}^i_a$ on the slice and with respect to the time direction $0$, we can then solve for $L^{ab}$,
\begin{equation}
    L^{ab} \propto \epsilon^{abc}    \hat{e}^i_c \epsilon_{uvw} X_i^{uvw}.
\end{equation}
Now, when $[ABC]$ does contain $0$, we have instead
\begin{equation}
    L^{0a} e_i^{b}-L^{0b} e_i^{a}+ L^{ab} e^0_i = X_i^{0ab};
\end{equation}
we have already solved for $L^{ab}$, so we can again dualise and contract with the inverse triad to arrive at
\begin{equation}
    L^{0a} \propto \epsilon^{abc}\hat{e}^i_b ( \epsilon_{cde} X^{0de}_i - \epsilon_{cde} L^{de} e^0_i ).
\end{equation}
This establishes injectivity, so that if there is a solution $L[\phi]$, then it is unique and given by the above expression. \\
However, there is a 6-dimensional cokernel to equation \ref{EqForL}, which constrains the connections $\omega$ for which $L$ exists. This cokernel is not 6D by coincidence - it is precisely tangential to the kernel of the presymplectic form of ECH gravity. Therefore, fixing the degeneracy as explained in appendix \ref{App:tech} also gives a way to fix the existence problem for $L$. 
In the notation of the literature\cite{cattaneoReducedPhaseSpace2019}, there is a projection map $p' = p'_{1,2} $ which we can apply to $\omega\wedge\phi$ to put it in the subspace where the cokernel is trivial. Then, the solution $L[\phi]$ always exists uniquely if we phrase the equation as \ref{EqForL}.


\section{Variation of $P_\phi$}\label{App:Var}
We vary the charge \ref{ShiftCharge} in pieces. We begin with the Einstein constraint bulk piece,
\begin{align}
    \Tr[\delta(-(\phi\wedge e)_\beta\wedge F_{ \omega})] &= 
    \delta e_I (F_{ \omega})^{IJ}_\beta \phi_J 
    + d(\Tr[(\phi\wedge e)_\beta\wedge\delta{ \omega}]) 
    - \Tr[d_{ \omega}(\phi\wedge e)_\beta\wedge\delta{ \omega}].
\end{align}
Next, we have the bulk term corresponding to the Gauss piece,
\begin{equation}
	\begin{aligned}
		\delta(-d_\omega e_I \wedge \omega_\beta^{IJ}\phi_J) =& -d_\omega\delta e_I\wedge \omega_\beta^{IJ}\phi_J 
		- e_I \wedge\delta\omega^{IJ}\wedge  p'((\omega_\beta)_{JK}\phi^K) 
		- d_\omega e_I \wedge \delta\omega_\beta^{IJ}\phi_J \\
		=& -d( \delta e_I\wedge \omega_\beta^{IJ}\phi_J  ) -\delta e_I\wedge d_\omega(\omega_\beta^{IJ}\phi_J)\\
		&+ e_I \wedge  (\omega_\beta)_{JK}\phi^K\wedge \delta\omega^{IJ}
		- d_\omega e_I \wedge \delta\omega_\beta^{IJ}\phi_J
	\end{aligned}
\end{equation}
We can now see in the first term in the last line that only $e\wedge (\omega_\beta\cdot\phi)$ appears, which automatically lets us rewrite 
\begin{equation}
	e\wedge (\omega_\beta\cdot\phi) = e\wedge p'(\omega_\beta\cdot\phi) = e\wedge (L[\phi]_\beta\cdot e) = -[\frac{1}{2}e^2,L[\phi]_\beta]
\end{equation}
and therefore
\begin{equation}
\begin{aligned}
		 e_I \wedge  (\omega_\beta)_{JK}\phi^K\wedge \delta\omega^{IJ} 
	&=-\Tr[[\frac{1}{2}e^2,L[\phi]_\beta]\wedge\delta\omega] \\
	&= -\Tr[[\frac{1}{2}e^2,L[\phi]]\wedge\delta\omega_\beta]  
	= \Tr[(e\wedge L[\phi]\cdot e) \wedge \delta\omega_\beta].
\end{aligned}
\end{equation}

We can then combine the bulk pieces to get 
\begin{equation}
	\begin{gathered}
		\delta e_I\wedge \left(
		(F_{ \omega})^{IJ}_\beta \phi_J - d_\omega(p'(\omega_\beta^{IJ}\phi_J))
		\right)  \\
		- \Tr[
		\left(
		d_{ \omega}(\phi\wedge e) - L[\phi]\cdot e\wedge e -\phi\wedge d_\omega e
		\right)
		\wedge\delta{ \omega}_\beta]
	\end{gathered}
\end{equation}

We can see that the $\phi\wedge d_{ \omega} e$ term in the second row cancels the piece which would lead to a torsion piece $\mathbb{T}_\phi$ in the transformation of the tetrads\cite{langenscheidtNewEdgeModes2025}. Further, there is now instead a torsion piece in the first row, which modifies the transformation law of the connection, which would otherwise contain $d_{ \omega} L[\phi]$, as usual for Lorentz transformations. In particular, actually only $p'(\omega_\beta\cdot\phi)$ appears. We also see that the curvature piece coming from the first term in the first line is unchanged with respect to the naive shifts.\\

Overall, we find that the charge generates the bulk vector field
\begin{equation}
    Y_\phi[ e] = d_{ \omega}\phi - L[\phi]\cdot e \quad Y_\phi[ \omega]\wedge e = d_\omega( L[\phi]\wedge e) - F_\omega\wedge\phi.
\end{equation}
\\
We now look at the corner pieces. Since the bulk pieces already define the vector field fully, the corner pieces must cancel and vanish. The full corner variation reads
\begin{equation}
    \begin{gathered}
        \Tr[(\phi\wedge e)_\beta\wedge\delta{ \omega} - (\delta e\wedge\phi)\wedge \omega_\beta ] - \delta p_I \phi^I\\
        =-\phi_I \delta \omega^{IJ}_\beta  e_J
        -\delta e_I   \omega^{IJ}_\beta\phi_J +\phi_I\delta \omega^{IJ}_\beta e_J - \delta e_J  \omega^{IJ}_\beta \phi_I = 0
    \end{gathered}
\end{equation}
and so we have
\begin{equation}
    \boxed{
    I_{Y_\phi}\Omega + \delta P_\phi =  0
    }
\end{equation}
which means the bulk vector field is perfectly integrable for all field-independent $\phi$.
The on-shell value of the charges is also nonzero for a generic configuration $( e, \omega)$ for all values of $\phi$. 



\section{Full bulk commutator}\label{App:Commutator}
Here, we calculate the actual algebra of charges. We omit the algebra of Lorentz charges as these are fairly standard in the literature\cite{geillerMostGeneralTheory2021,freidelEdgeModesGravity2020}.
\subsection{Lorentz-Shift}
    We have for the bulk pieces
    \begin{equation}
        \Tr[X_\alpha[(\phi\wedge e)_\beta\wedge F_\omega] ] = \Tr[(\alpha \cdot\phi\wedge  e)_\beta\wedge F_\omega]
    \end{equation}
    and
    \begin{equation}
        \Tr[X_\alpha[\frac{1}{2}d_\omega e^2\, L[\phi]] ] =\Tr[\frac{1}{2}d_\omega e^2\,[\alpha, L[\phi]] + \frac{1}{2}d_\omega e^2\, X_\alpha[ L[\phi]] ]
    \end{equation}
    so we have in the bulk
    \begin{equation}
        \{J_\alpha,P_\phi\} = P_{\alpha\cdot\phi} + J_{ A(\alpha,\phi)}
    \end{equation}
    with
    \begin{equation}
        A(\alpha,\phi):= X_\alpha[ L[\phi]]  - L[\alpha\cdot\phi] - [ L[\phi],\alpha]
    \end{equation}
    a piece that measures the Lorentz anomaly of $L[-]$: The action of Lorentz transformations should affect it at least through its argument and through commutators. However, it carries an anomaly on top of this. This is sensible from the definition of $ L[\phi]$, as all pieces except $\omega$ may be acted upon tensorially by the Lorentz transformations. If not for the total derivative term, we would have $A=0$. So, we actually have
    \begin{equation}
        A(\alpha,\phi)\wedge e = d\alpha\wedge\phi
    \end{equation}
    and the additional corner piece is a reflection of the fact that $A\neq 0$, or equivalently that $L[-]$ transforms noncovariantly under Lorentz transformations.\\
    With corners, we have
    \begin{equation}
        \{J_\alpha,P_\phi\}  \approx- \oint \Tr[(p\wedge\phi)\,\alpha + d\alpha_\beta \wedge (\phi\wedge e)]
    \end{equation}
    This is equal to the corner term of $P_{\alpha\cdot\phi} + J_{ A(\alpha,\phi)}$ if we identify
    \begin{equation}
        - \phi\cdot d\alpha_\beta  =  e\cdot A(\alpha,\phi)_\beta \Leftrightarrow  A(\alpha,\phi)\wedge  e =   d\alpha \wedge\phi .
    \end{equation}
    So the Lorentz-anomaly of $L[-]$ is what gives rise to the extra term.

    Let us compare this to the abstract model algebra; there, we would have the field-dependent Lorentz transformations $\alpha,\Tilde{\alpha}$ and field-independent translations $\phi,\Tilde{\phi}$ in the shape
    \begin{equation}
        [(\alpha,\phi),(\Tilde{\alpha},\Tilde{\phi})] = ( [\Tilde{\alpha},\alpha] - \delta_\alpha\Tilde{\alpha} + \delta_{\Tilde{\alpha}}\alpha , -\alpha\cdot\Tilde{\phi} + \Tilde{\alpha}\cdot\phi  ).
    \end{equation}
    If we specialise this to $\phi=0,\Tilde{\alpha}= L[\phi]$ and $\alpha$ field-independent, then
    \begin{equation}
        [(\alpha,0),( L[\tilde{\phi}],\Tilde{\phi})] = ( [ L[\phi],\alpha] - X_\alpha [L[\phi]]  , -\alpha\cdot\Tilde{\phi}  ) = (-A(\alpha,\phi) + \alpha_{-\alpha\cdot\Tilde{\phi}}, -\alpha\cdot\Tilde{\phi} )
    \end{equation}
    So that as expected, we have a perfect representation of this algebroid both in bulk and corners. There are no central terms at this level. It is also possible to calculate the commutator of transformations explicitly, and the result matches this algebroid structure. We do not show this here, as the integrability of transformations guarantees that they match.
    
\subsection{Shift-Shift}
    The calculation for the shift-shift Poisson brackets are much more involved, but still doable in practice, unlike the direct computation of the commutators of transformations, which contain too many implicit terms. 
    We split the calculation into two pieces, one for the Einstein term, one for the Gauss term. We will omit the use of traces $\Tr[-]$ in most steps and make them explicit whereever necessary.\\
    We will use
    \begin{align}
        Y_\phi[ e^2] = 2 d_\omega(\phi\wedge e) + [ e^2, L[\phi]] - 2\phi\wedge d_\omega e
    \end{align}
    and
    \begin{align}
        Y_\phi[(F_\omega)]_\beta \wedge e &= d_\omega (Y_\phi[\omega]_\beta \wedge e)+Y_\phi[\omega]_\beta\wedge d_\omega e \\
        &= d_\omega ( d_\omega \alpha_\beta \wedge e + \alpha_\beta \wedge d_\omega e)
        -d_\omega (F_\beta\wedge \phi) +Y_\phi[\omega]_\beta\wedge d_\omega e\\
        &= 
        d_\omega^2 \alpha_\beta \wedge e  + \alpha_\beta \cdot d_\omega^2 e
        -F_\beta\cdot d_\omega \phi +Y_\phi[\omega]_\beta\wedge d_\omega e
    \end{align}
    as well as
    \begin{equation}
        [\phi\wedge e,\alpha] =  (-\alpha\cdot\phi) \wedge e + \phi\wedge(-\alpha\cdot e)
    \end{equation}
    Let us look first at the Einstein term.
    \begin{align}\label{Der1}
        Y_\phi[(\Tilde{\phi}\wedge e)_\beta \wedge F_\omega] &= 
        (\Tilde{\phi}\wedge d_\omega\phi)_\beta \wedge F_\omega
        + [(\Tilde{\phi}\wedge  e)_\beta, L[\phi]]\wedge F_\omega \\
        &+ ( L[\phi]\cdot\Tilde{\phi}\wedge  e)_\beta \wedge F_\omega + (\Tilde{\phi}\wedge e) \wedge  Y_\phi[(F_\omega)]_\beta 
    \end{align}
    We examine the last term:
    \begin{align}
        &\Tr[ (\Tilde{\phi}\wedge e) \wedge  Y_\phi[(F_\omega)]_\beta ]=\Tilde{\phi}\cdot Y_\phi[(F_\omega)]_\beta \cdot e \\
        &=\Tr[(\Tilde{\phi}\wedge e)_\beta\wedge [F_\omega, L[\phi]]   - (\Tilde{\phi}\wedge d_\omega \phi)_\beta\wedge F_\omega ]\\
        &+ \Tr[(\Tilde{\phi}\wedge d_\omega^2 e)\, \alpha_\beta+ (\Tilde{\phi}\wedge d_\omega e)\wedge Y_\phi[\omega]_\beta  ]
    \end{align}
    The first two terms cancel with the first two in \ref{Der1}, so 
    
    \begin{align}\label{Der2}
        Y_\phi[(\Tilde{\phi}\wedge e)_\beta\wedge F_\omega] &=( L[\phi]\cdot\Tilde{\phi}\wedge  e)_\beta\wedge F_\omega + (\Tilde{\phi}\wedge d_\omega^2 e)_\beta\wedge  L[\phi]+ (\Tilde{\phi}\wedge d_\omega e)\wedge Y_\phi[\omega]_\beta 
    \end{align}
    where we already see that in the first term, we have the action $\Tilde{\phi}\mapsto  L[\phi] \cdot \Tilde{\phi}$. The other terms only involve torsion and are therefore functions of the Gauss constraint.\\
    As for the Gauss piece itself, we can analyse it using its explicit form
    \begin{equation}
        \Tr[\frac{1}{2}d_\omega e^2_\beta\cdot  L[\tilde{\phi}] ]= d_\omega e \cdot \omega_\beta \cdot \Tilde{\phi}
        == d_\omega e \cdot p'(\omega_\beta \cdot \Tilde{\phi})
    \end{equation}
    which we can act on with $Y_\phi$ directly to produce
    \begin{equation}\label{Der3}
        \begin{aligned}
            \Tr[ Y_\phi[\frac{1}{2}d_\omega e^2_\beta\cdot  L[\tilde{\phi}]]] &= \Tr[(d_\omega e\wedge\Tilde{\phi})_\beta\wedge Y_{\phi}[\omega]  ]\\
            &+ d_\omega Y_\phi[ e]_I \, ( L[\tilde{\phi}])_\beta^{IJ} \wedge  e_J + (Y_\phi[\omega]\cdot e)_I\wedge ( L[\tilde{\phi}])_\beta^{IJ} \cdot  e_J.
        \end{aligned}
    \end{equation}
    We can rewrite the third term as
    \begin{equation}
        (Y_\phi[\omega]\cdot e)_I \wedge( L[\tilde{\phi}])_\beta^{IJ} \cdot  e_J = \Tr[\frac{1}{2}[Y_\phi[\omega], e^2]_\beta\, L[\tilde{\phi}]].
    \end{equation}
     Now the first piece of \ref{Der3} cancels with the last piece of \ref{Der2}, yielding
\begin{equation}
    \begin{aligned}
         Y_\phi[(\Tilde{\phi}\wedge e)_\beta F_\omega+\frac{1}{2}d_\omega e^2\wedge  L[\tilde{\phi}]]  
        &=( L[\phi]\cdot\Tilde{\phi}\wedge  e)_\beta\wedge F_\omega + (\Tilde{\phi}\wedge d_\omega^2 e)_\beta\wedge  L[\phi]\\
        &+ \frac{1}{2}[Y_\phi[\omega], e^2]_\beta\wedge L[\tilde{\phi}]
        +
        (d_\omega Y_\phi[ e]\wedge e)\wedge  ( L[\tilde{\phi}])_\beta .
    \end{aligned}
\end{equation}
To write this in a manifestly antisymmetric form, we also need contributions coming from the boundary term. To ease notation, we introduce for the following the shorthand
\begin{equation}
    Q^I_\phi = ( L[\phi])^{IJ}_\beta e_J = p'((\omega)^{IJ}_\beta \phi_J).
\end{equation}
We have
\begin{equation}
\begin{gathered}
        Y_\phi[p_I \Tilde{\phi}^I] = - \Tilde{\phi}_I ( d_\omega Q_\phi^I - (F_\omega)^{IJ}_\beta \phi_J ) - \Tilde{\phi}_I \omega^{IJ}_\beta Y_\phi[ e]\\
        =  - \Tr[(\phi\wedge\Tilde{\phi})\, (F_\omega)_\beta] - \Tilde{\phi}_I  d_\omega Q_\phi^I  + Q_{\Tilde{\phi}}^I \wedge Y_\phi[ e]_I
\end{gathered}
\end{equation}
of which we carry over the second and third term into the bulk,
\begin{equation}
    \begin{aligned}
        d(- \Tilde{\phi}_I  d_\omega Q_\phi^I  + Q_{\Tilde{\phi}}^I \wedge Y_\phi[ e]_I) &= d_\omega Q_{\Tilde{\phi}}^I \wedge Y_\phi[ e]_I - Q_{\Tilde{\phi}}^I \wedge d_\omega Y_\phi[ e]_I\\
        &- d_\omega \Tilde{\phi}_I  d_\omega Q_\phi^I - \Tilde{\phi}_I  d_\omega^2 Q_\phi^I
    \end{aligned}
\end{equation}
and we rewrite
    \begin{equation}
    \begin{gathered}
                \Tr[\frac{1}{2}[Y_\phi[\omega], e^2]_\beta\, L[\tilde{\phi}]]
        = 
        \Tr[(\phi\wedge  L[\tilde{\phi}]\cdot  e)_\beta\wedge F_\omega] + d_\omega Q_\phi^I \wedge ( L[\tilde{\phi}]\cdot  e)_I\\
        =
        \Tr[(\phi\wedge e)_\beta\,[F_\omega, L[\tilde{\phi}]]
        -( L[\tilde{\phi}]\cdot\phi\wedge   e)_\beta\wedge F_\omega]
        + d_\omega Q_\phi^I \wedge ( L[\tilde{\phi}]\cdot  e)_I
    \end{gathered}
    \end{equation}
    and furthermore 
    \begin{equation}
        \begin{gathered}
        \Tr[(\phi\wedge e)_\beta\wedge[F_\omega, L[\tilde{\phi}]] ]=
        \Tr[[(\phi\wedge e),F_\omega]_\beta\, L[\tilde{\phi}]]\\ 
        =
        \Tr[(
         e\wedge (F_\omega\cdot\phi)
        -\phi\wedge (F_\omega\cdot e)
        )_\beta\; L[\tilde{\phi}]]\\
        = 
        -Q_{\Tilde{\phi}}^I\wedge d_\omega^2\phi_I
        -
        \Tr[(
        \phi\wedge d_\omega^2 e
        )_\beta\, L[\tilde{\phi}]].
        \end{gathered}
    \end{equation}
    Now we can combine the pieces into
    \begin{equation}
    \begin{gathered}
        \Tr[ Y_\phi[(\Tilde{\phi}\wedge e)_\beta\wedge F_\omega+\frac{1}{2}d_\omega e^2\,  L[\tilde{\phi}]] ]\\
        =\Tr[( L[\phi]\cdot\Tilde{\phi}\wedge  e)_\beta \wedge F_\omega 
        + (\Tilde{\phi}\wedge d_\omega^2 e)_\beta\,  L[\phi]
        +
        (d_\omega Y_\phi[ e]\wedge e) \,  L[\tilde{\phi}]_\beta  \\
        +(\phi\wedge e)_\beta\wedge [F_\omega, L[\tilde{\phi}]]
        -( L[\tilde{\phi}]\cdot\phi\wedge   e)_\beta\wedge F_\omega ]
        + d_\omega Q_\phi^I \wedge ( L[\tilde{\phi}]\cdot  e)_I\\
        +d_\omega Q_{\Tilde{\phi}}^I \wedge Y_\phi[ e]_I 
        - Q_{\Tilde{\phi}}^I \wedge d_\omega Y_\phi[ e]_I
        - d_\omega \Tilde{\phi}_I \wedge d_\omega Q_\phi^I 
        - \Tilde{\phi}_I  d_\omega^2 Q_\phi^I\\
        = 
        \Tr[(( L[\phi]\cdot\Tilde{\phi}- L[\tilde{\phi}]\cdot\phi)\wedge  e)_\beta\wedge F_\omega 
        -
        (
        \phi\wedge d_\omega^2 e
        )_\beta\, L[\tilde{\phi}] 
        + (\Tilde{\phi}\wedge d_\omega^2 e)_\beta\, L[\phi] ]
        \\
        - d_\omega Q_\phi^I \wedge Y_{\Tilde{\phi}}[ e]_I  
        +d_\omega Q_{\Tilde{\phi}}^I \wedge Y_\phi[ e]_I 
        -Q_{\Tilde{\phi}}\cdot d_\omega^2\phi
        - \Tilde{\phi}_I  d_\omega^2 Q_\phi^I
    \end{gathered}
\end{equation}
We can rewrite the last two terms as
\begin{equation}
    -Q_{\Tilde{\phi}}^I\wedge d_\omega^2\phi_I
        - \Tilde{\phi}_I  d_\omega^2 Q_\phi^I\\
        =
        d(Q_{\Tilde{\phi}}^I\wedge d_\omega\phi_I-\Tilde{\phi}_I  d_\omega Q_\phi^I)
        -d_\omega Q_{\Tilde{\phi}}^I\cdot d_\omega\phi_I
        + d_\omega\Tilde{\phi}_I  d_\omega Q_\phi^I
\end{equation}
and
\begin{equation}
    Q_{\Tilde{\phi}}^I\wedge d_\omega\phi_I-\Tilde{\phi}_I  d_\omega Q_\phi^I
    =
    -d(  Q_{\Tilde{\phi}}^I \phi_I) + \phi_I d_\omega Q_{\Tilde{\phi}}^I
    -\Tilde{\phi}_I  d_\omega Q_\phi^I
\end{equation}
So that everything is manifestly antisymmetric\footnote{We do not need the codimension 3 term, but we can even write this one as the antisymmetric $\phi_I  Q_{\Tilde{\phi}}^I = \Tr[\omega_\beta\cdot (\phi\wedge\Tilde{\phi}  )]$.} in $\phi,\Tilde{\phi}$. This is as expected since we are dealing with Hamiltonian charges, however it is, as demonstrated, a (very) nontrivial consistency check. The bulk now consists of
\begin{equation}
    \begin{gathered}
        \Tr[(( L[\phi]\cdot\Tilde{\phi}- L[\tilde{\phi}]\cdot\phi)\wedge  e)_\beta F_\omega 
        -
        (
        \phi\wedge d_\omega^2 e
        )_\beta\wedge L[\tilde{\phi}]
        + (\Tilde{\phi}\wedge d_\omega^2 e)_\beta\wedge  L[\phi]]\\
        + d_\omega Q_\phi^I \wedge ( L[\tilde{\phi}]\cdot e)_I 
        -d_\omega Q_{\Tilde{\phi}}^I \wedge ( L[\phi]\cdot e)_I ,
    \end{gathered}
\end{equation}
and the boundary piece consists of
\begin{equation}
    \Tr[- (\phi\wedge\Tilde{\phi})\, (F_\omega)_\beta ]+\phi_I d_\omega Q_{\Tilde{\phi}}^I
    -\Tilde{\phi}_I  d_\omega Q_\phi^I
\end{equation}
However, a meaningful bulk-boundary split only happens when we go on-shell of the bulk constraints. We therefore impose $d_\omega e=0=(F_\omega)_\beta\cdot e$
which gives
\begin{equation}
    d_\omega Q_\phi = (d_\omega  L[\phi])_\beta\cdot e\quad d_\omega^2 Q_\phi = [F_\omega , L[\phi]]_\beta\cdot e
\end{equation}
then in the bulk we are left with
\begin{equation}
     e_I\wedge (d_\omega ( L[\phi])_\beta \cdot  L[\tilde{\phi}]
     -d_\omega ( L[\tilde{\phi}])_\beta \cdot  L[\phi])^{IJ} \wedge e_J  
\end{equation}
while on the boundary, we have
\begin{equation}
    - \Tr[(\phi\wedge\Tilde{\phi})\, (F_\omega)_\beta ]+\phi_I (d_\omega  L[\tilde{\phi}])_\beta^{IJ}\wedge e_J
    -\Tilde{\phi}_I  (d_\omega  L[\phi])_\beta^{IJ}\wedge e_J.
\end{equation}
We now study the bulk piece by comparing it to
\begin{equation}
    d_\omega[ L[\phi], L[\tilde{\phi}]]_\beta.
\end{equation}
To do so, we use a number of identities of traces and the homomorphism $(-)_\beta$. We focus in this first on the first term of the two. First, we rewrite it in the manifestly cyclic form\footnote{We can also just do the classic $AB = \frac{1}{2}[A,B]+ \frac{1}{2}\{A,B\}$ on $A= (d_\omega L[\phi])_\beta, B=  L[\tilde{\phi}]$.}(see appendix \ref{App:Id})
\begin{equation}
     \begin{aligned}
         e_I\wedge (d_\omega ( L[\phi])_\beta \cdot  L[\tilde{\phi}])^{IJ}\wedge e_J 
         &= ( e^2)^I_{\; J} \wedge (d_\omega( L[\phi])_\beta)^J_{\;K} ( L[\tilde{\phi}])^K_{\; I} \\
         &= \text{Tr}[ e^2\wedge d_\omega( L[\phi])_\beta  L[\tilde{\phi}]]
     \end{aligned}
\end{equation}
Then, we can rewrite this as 
\begin{equation}
    \Tr[-\frac{1}{2} e^2 \wedge [d_\omega( L[\phi])_\beta,  L[\tilde{\phi}]]] =\Tr[ -\frac{1}{2} e^2_\beta \wedge [d_\omega L[\phi],  L[\tilde{\phi}]]]
\end{equation}
and together with the antisymmetric counterpart, we have in the bulk
\begin{equation}
    \Tr[\,-\frac{1}{2} e^2_\beta \wedge d_\omega[ L[\phi],  L[\tilde{\phi}]]\,] \approx \Tr[d( -\frac{1}{2} e^2_\beta \wedge [ L[\phi],  L[\tilde{\phi}]] )]
\end{equation}
so it becomes a pure boundary term. Therefore, we have on-shell:
\begin{equation}
\begin{aligned}
        \{P_\phi,P_{\Tilde{\phi}}\} &\approx \Tr[\,(\phi\wedge\Tilde{\phi})_\beta\, F_\omega 
    +\frac{1}{2} e^2_\beta \, [ L[\phi],  L[\tilde{\phi}]] \, ] \\
    &-\phi_I\cdot d_\omega( \omega_\beta\cdot \Tilde{\phi})^I
    +\Tilde{\phi}_I\cdot  d_\omega (\omega_\beta\cdot {\phi})^I
    \end{aligned}
\end{equation}
Then, through carefully expanding out the second and third term, and using again the trace-commutator identities, we reach the final result
\begin{equation}
    \begin{aligned}
    \{P_\phi,P_{\Tilde{\phi}}\} &\approx \oint_{\partial\Sigma}\Tr[\,-(\phi\wedge\Tilde{\phi})_\beta\, F_\omega +\frac{1}{2} e^2_\beta \, [ L[\phi],  L[\tilde{\phi}]] + \omega\wedge d(\phi\wedge\Tilde{\phi})_\beta \, ]  \\
    &=\oint_{\partial\Sigma} q_{\phi,\Tilde{\phi}}
    \end{aligned}
\end{equation}
which contain a Bianchi piece, a Lorentz piece and a nontrivial extension which vanishes for constant parameters.\\
Off-shell, the situation is more complicated, but by suitable further manipulations in the bulk we find
\begin{equation}\label{OffshellBracket}
\boxed{
\begin{gathered}
        \{P_\phi,P_{\Tilde{\phi}}\} = 
    \int_\Sigma 
    \Tr[\,-(( L[\phi]\cdot\Tilde{\phi}- L[\tilde{\phi}]\cdot\phi)\wedge  e)_\beta \wedge F_\omega 
    -\frac{1}{2}d_\omega e^2_\beta\, 2[ L[\phi],  L[\tilde{\phi}]]\,]
    \\
        +\int_\Sigma \Tr[\,
        (
        \phi\wedge d_\omega^2 e
        )_\beta\, L[\tilde{\phi}]
        - (\Tilde{\phi}\wedge d_\omega^2 e)_\beta\,  L[\phi] \,]
    +
    \oint_{\partial\Sigma} q_{\phi,\Tilde{\phi}}
\end{gathered}
}
\end{equation}
Now this is the kind of result we would expect; the shifts here contain Lorentz transformations, therefore it is reasonable that the translations can act nontrivially on each other. This is not just an artifact of the bulk parametrisation, but instead persists to the onshell phase space as seen above. We can see an Einstein constraint piece with a Lorentz-on-translation action, as well as commutators of the Lorentz parameters $L$. A curious addition is the terms involving $d^2_\omega e$, which are technically speaking also just Gauss constraint terms\footnote{ They are also reminiscent of the \say{edge simplicity constraints}\cite{ashtekarQuantumTheoryGeometry1998} $F_\omega\cdot e$, which are implied in the continuum, but not so in discretizations. Their presence in the continuum may indicate subtle structures in lattice applications.}. 

\section{Extended BF theory}\label{App:eBF}
Here we introduce an extension of BF theory that, in hindsight, is quite natural as it leads to a symmetry structure that is quite similar to the one of tetrad gravity, even before constraining. If we start from a Poincaré-BF theory with 1-form fields $\omega^{IJ},\theta^I$ and 2-form fields $B^{IJ},t^I$, we can arrive at tetrad gravity by imposing the simplicity constraint
\begin{equation}
    B=\frac{1}{2}e^2_\beta
\end{equation}
through a Lagrange multiplier 2-form $\lambda^{IJ}$:
\begin{equation}
    L_0 = t_I\wedge d_\omega e^I +  \Tr[ B\wedge F_\omega -\lambda \wedge ( 
B-\frac{1}{2} e^2_\beta )]
\end{equation}
This reduces to ECH gravity after eliminating $\lambda$ and setting $t=0$, which is a consequence of the equations of motion. The resulting phase space does not contain $\lambda$, which makes it impossible to impose the equations of motion on the phase space level, which involve $\lambda$. So, we can add a kinetic term for it with an additional 1-form \say{dual connection} $\pi^{IJ}$,
\begin{equation}
    L_0 = t_I\wedge d_\omega e^I +  \Tr[ B\wedge F_\omega + \lambda\wedge d_\omega\pi -\lambda \wedge ( 
B-\frac{1}{2} e^2_\beta )]
\end{equation}
which technically is just the result of applying a Kalb-Ramond shift $B\mapsto B+d_\omega\pi$ to the former Lagrangian.
This comes with associated symplectic potential
\begin{equation}
     \theta_1 = -t_I\wedge \delta e^I +  \Tr[ B\wedge \delta\omega + \lambda\wedge \delta\pi]
\end{equation}
and equations of motion
\begin{equation}
   \begin{gathered}
       E_0= \delta t_I\wedge d_\omega e^I + (d_\omega t^I+  e_J \lambda^{JI}_\beta)\wedge\delta e_I\\
       +\Tr [
   \delta B \wedge (F_\omega-\lambda) +\delta\lambda\wedge(d_\omega\pi -(B-\frac{1}{2} e^2_\beta)) \\
   - d_\omega\lambda\wedge\delta\pi - (d_\omega B -  e\wedge t+ [\pi,\lambda])\wedge\delta\omega
   ].
   \end{gathered}
\end{equation}

The EoM of $B$ determines $\lambda$ through the curvature constraint and returns us to the ECH phase space, up to a corner term. This is true even though the equation of motion for $\lambda$, which is the (modified) simplicity constraint,
\begin{equation}
    B= \frac{1}{2}e^2_\beta + d_\omega\pi,
\end{equation}
is not exactly setting $B$ to be the gravitational flux, but instead relates the failure of the simplicity constraints to the curvature $d_\omega\pi$ of $\pi$. Explicitly, we can rewrite
\begin{equation}
\begin{gathered}
        B\wedge\delta\omega = (B-\frac{1}{2} e^2_\beta - d_\omega\pi)\delta\omega
    +\frac{1}{2} e^2_\beta \delta\omega + d_\omega\pi\delta\omega\\
    =\frac{1}{2} e^2_\beta \delta\omega
    + d(\pi\delta\omega) + \pi \delta F_\omega + (B-\frac{1}{2} e^2_\beta - d_\omega\pi)\delta\omega
\end{gathered}
\end{equation}
The last term vanishes when the modified simplicity constraints hold, and the third combines with the term of the $\lambda,\pi$ pair:
\begin{equation}
    \pi \delta F_\omega + \lambda\delta\pi = (\lambda-F_\omega)\delta\pi + \delta(\pi F_\omega)
\end{equation}
The first term again vanishes on-shell of the curvature constraint. We therefore return to the symplectic potential
\begin{equation}
     \theta_1 = \frac{1}{2} e^2_\beta \delta\omega
    -t_I\wedge \delta e^I
    + d(\pi\delta\omega) 
    + \delta(\pi F_\omega)
\end{equation}
and have removed any trace of $\pi,\lambda$ from the bulk, and $\pi$ appears only as a conjugate variable to $\omega$ on the corner. We can, if we want, always remove these corner pieces by adding a boundary Lagrangian or exploiting corner ambiguities. \\
The first two pieces are GR in the 'usual' way: The symplectic form 
\begin{equation}
    \delta e\wedge(\delta t - \delta\omega_\beta \wedge  e)
\end{equation}
has a kernel $V_\Delta = \Delta \frac{\delta}{\delta \omega} + \Delta_\beta\wedge e\frac{\delta}{\delta t}$ parametrised by Lie algebra valued 1-forms $\Delta^{IJ}$. The correct way to gauge fix this is to set $t=0$, and then impose the structural constraints as in \ref{App:tech}. This then brings one exactly to the phase space of GR. Alternatively, we can set $F_\omega=0$, which completely removes the kernel up to Lorentz transformations. However, this is only an option on the level of the symplectic form, and prohibited by the dynamics, which selects $t=0$ and the structural constraints as necessary for a consistent dynamics\footnote{This can be determined through applying the Dirac algorithm.}.\\
Each equation of motion gives us a generator of gauge transformations. The ones for $\omega, e,\pi$ (the 'connections') give us

\begin{equation}  
    J_\alpha = \int \Tr[ B\wedge d_\omega\alpha + ( e\wedge t)\,\alpha + [\lambda,\pi]\,\alpha ]
\end{equation}
\begin{equation}
    T_\phi = -\int t_I\wedge d_\omega\phi^I + \Tr[(\phi\wedge e)_\beta {\wedge}\lambda]
\end{equation}
\begin{equation}
    R_\beta = \int \Tr[\lambda\wedge d_\omega\beta ]
\end{equation}
which are the Lorentz, shift and Bianchi generators\footnote{We call this a Bianchi generator because on-shell of $\lambda=F_\omega$, the bulk constraint is the Bianchi identity $d_\omega F_\omega=0$ and therefore automatically zero.}.
The other three equations of motion, for $B,t,\lambda$, give us
\begin{equation}
    K_\mu = -\int \Tr[\mu\wedge(F_\omega - \lambda)] -\oint \Tr[\mu\wedge\omega]
\end{equation}
\begin{equation}
    \tau_v = \int  e_I\wedge d_\omega v^I
\end{equation}
\begin{equation}
    MSC_\Delta = -\int \Tr[\pi\wedge d_\omega\Delta - (B-\frac{1}{2} e^2_\beta)\wedge\Delta]
\end{equation}
which are the curvature, torsion and modified simplicity generators. \\
It should be clear that $T_\phi$ is the closest pendent to the shift generators. In particular, the $\lambda$-piece becomes the Einstein constraint once $K=0$ for bulk parameters. It is important to note that all of these functions are \textit{generators}, but they are only \textit{constraints} when their parameters have no support on the corner.\\

These gauge generators have a quite intriguing algebra. The first three of them fulfil
\begin{equation}
    \{J_\alpha, T_\phi\}=T_{\alpha\cdot\phi},\{J_\alpha,R_\beta\}=R_{[\alpha,\beta]},\{J_\alpha,J_{\tilde{\alpha}}\}=J_{[\alpha,\tilde{\alpha}]}
\end{equation}
and 
\begin{equation}
    \{T_\phi,T_\psi\} = R_{(\phi\wedge\psi)_\beta}, 
    \{T,R\}=\{R,R\}=0
\end{equation}
which is a central extension of the Poincare algebra ($J,T$) by an abelian $\mathfrak{so}(1,3)^\ast$, represented by $R$. This has the structure of a double extension
\begin{equation}
    \mathfrak{h}\ltimes (\mathfrak{g}\times_c \mathfrak{h}^\ast)
\end{equation}
with $\gfrak=\rbb^{1,3}, \mathfrak{h}=\frak{so}(1,3) $, which is part of the largest class of Lie algebras admitting invariant, nondegenerate bilinear forms\cite{figueroa-ofarrillNonsemisimpleSugawaraConstructions1994}. This particular algebra is also known under the name of the 'Maxwell algebra' and is known to describe the symmetries of a particle in a fixed electromagnetic background\cite{salgadoTopologicalGravityTransgression2014,bacryGrouptheoreticalAnalysisElementary1970,schraderMaxwellGroupQuantum1972,azcarragaGeneralizedCosmologicalTerm2011}.\\
The other algebra is more subtle,
\begin{equation}
    \{MSC_\Delta, K_\mu\}=-\oint_{\p\s} \Tr[\mu\wedge\Delta], \{MSC_\Delta, \tau_v\} = \int_{\s}  e_I\wedge\Delta^{IJ}\wedge v_J
\end{equation}
\begin{equation}
    \{MSC,MSC\}=\{K,K\}=\{\tau,\tau\}=\{K,\tau\}=0
\end{equation}
in which we can see that the \textit{constraints} $MSC,K$ commute (bulk-supported parameters), whereas their corner pieces $\pi,\omega$ fulfil a canonical commutation relation. So on the corner, $\pi,\omega$ are canonically conjugate. However, $\tau$ does not behave well with the other constraints. This is in principle okay, as it is a redundant constraint (implied by $d_\omega e^2=0$).\\
Finally, we have the Poisson brackets between the two sets (without $J$, which just rotates parameters):
\begin{equation}
    \begin{gathered}
        \{MSC_\Delta,T_\phi\}= \tau_{\Delta_\beta\cdot\phi} - \int t_I\wedge \Delta^{IJ}\phi_J, \\
        \{MSC_\Delta,R_\beta\} = K_{[\Delta,\beta]} + \oint d\Delta\cdot\alpha\\
        \{T_\phi , \tau_v \}= \int d_\omega\phi_I\wedge d_\omega v^I\\
        \{T_\phi , K_\mu \}=\{R_\beta,K_\mu\}= 0 
    \end{gathered}
\end{equation}
So it is not first class in total. 
To get to ECH, we would need to impose $MSC=0=K$ in the bulk. This works, but leaves us with a presymplectic form. One might then naively conclude that one can actually set $F_\omega=0$ as a gauge fixing, but this is not compatible with the dynamics:\\

 As it turns out, the naive Hamiltonian is a weighted sum of the constraints,
\begin{equation}
    H = J_{\omega_t}-T_{ e_t}+R_{\pi_t}+MSC_{\lambda_t}+K_{B_t}+\tau_{t_t}
\end{equation}
so the Dirac algorithm amounts to asking the constraints to form a first class set. We have seen that they do not; so we need to take a look at the second class part. 
\begin{equation}
    \{MSC_{\lambda_t},T_{\phi}\}= \tau_{(\lambda_t)_\beta\cdot{\phi}} - \int t_I\wedge (\lambda_t)^{IJ}\phi_J
\end{equation}
\begin{equation}
    \{T_{ e_t},MSC_{\Delta},\}= \tau_{\dots} - \int t_I\wedge (\Delta)^{IJ}( e_t)_J
\end{equation}
This tells us that we have a bifurcating Dirac algorithm: If $t=0$, then the constraints are preserved as-is. Otherwise, we need to choose restrictions on the time evolution parameters $\lambda_t, e_t$ such that the second terms in the above vanish.\\
So, if we add $t=0$ to our constraints, we then need to require it, too is stabilised, and the algorithm continues. However, by analysing the symplectic form, we can see that setting
\begin{equation}
    t=0, MSC=0, K=0,
\end{equation}
together with the structural constraints, is enough to bring us back to ECH gravity. 

\bibliography{MyLibrary}

\begin{thebibliography}{30}
\providecommand{\natexlab}[1]{#1}
\providecommand{\url}[1]{\texttt{#1}}
\expandafter\ifx\csname urlstyle\endcsname\relax
  \providecommand{\doi}[1]{doi: #1}\else
  \providecommand{\doi}{doi: \begingroup \urlstyle{rm}\Url}\fi

\bibitem[Langenscheidt and Oriti(2025)]{langenscheidtNewEdgeModes2025}
Simon Langenscheidt and Daniele Oriti.
\newblock New edge modes and corner charges for first-order symmetries of {{4D}} gravity.
\newblock \emph{Classical and Quantum Gravity}, 42\penalty0 (7):\penalty0 075010, March 2025.
\newblock ISSN 0264-9381.
\newblock \doi{10.1088/1361-6382/adbfee}.

\bibitem[Cattaneo and Schiavina(2019)]{cattaneoReducedPhaseSpace2019}
Alberto~S. Cattaneo and Michele Schiavina.
\newblock The {{Reduced Phase Space}} of {{Palatini}}--{{Cartan}}--{{Holst Theory}}.
\newblock \emph{Annales Henri Poincar{\'e}}, 20\penalty0 (2):\penalty0 445--480, February 2019.
\newblock ISSN 1424-0661.
\newblock \doi{10.1007/s00023-018-0733-z}.

\bibitem[Horowitz(1991)]{horowitzTopologyChangeGeneral1991}
Gary~T. Horowitz.
\newblock Topology {{Change}} in {{General Relativity}}, September 1991.

\bibitem[Montesinos et~al.(2018)Montesinos, Gonzalez, and Celada]{montesinosGaugeSymmetriesFirstorder2018}
Merced Montesinos, Diego Gonzalez, and Mariano Celada.
\newblock The gauge symmetries of first-order general relativity with matter fields.
\newblock \emph{Classical and Quantum Gravity}, 35\penalty0 (20):\penalty0 205005, September 2018.
\newblock ISSN 0264-9381.
\newblock \doi{10.1088/1361-6382/aae10d}.

\bibitem[Geiller(2017)]{geillerEdgeModesCorner2017}
Marc Geiller.
\newblock Edge modes and corner ambiguities in 3d {{Chern}}--{{Simons}} theory and gravity.
\newblock \emph{Nuclear Physics B}, 924:\penalty0 312--365, November 2017.
\newblock ISSN 0550-3213.
\newblock \doi{10.1016/j.nuclphysb.2017.09.010}.

\bibitem[Geiller et~al.(2022)Geiller, Girelli, Goeller, and Tsimiklis]{geillerDiffeomorphismsQuadraticCharges2022}
Marc Geiller, Florian Girelli, Christophe Goeller, and Panagiotis Tsimiklis.
\newblock Diffeomorphisms as quadratic charges in 4d {{BF}} theory and related {{TQFTs}}, October 2022.

\bibitem[Canepa et~al.(2021)Canepa, Cattaneo, and Tecchiolli]{canepaGravitationalConstraintsLightlike2021}
Giovanni Canepa, Alberto~S. Cattaneo, and Manuel Tecchiolli.
\newblock Gravitational {{Constraints}} on a {{Lightlike}} boundary.
\newblock \emph{Annales Henri Poincar{\'e}}, 22\penalty0 (9):\penalty0 3149--3198, September 2021.
\newblock ISSN 1424-0637, 1424-0661.
\newblock \doi{10.1007/s00023-021-01038-z}.

\bibitem[Freidel et~al.(2017)Freidel, Perez, and Pranzetti]{freidelLoopGravityString2017}
Laurent Freidel, Alejandro Perez, and Daniele Pranzetti.
\newblock The loop gravity string.
\newblock \emph{Physical Review D}, 95\penalty0 (10):\penalty0 106002, May 2017.
\newblock ISSN 2470-0010, 2470-0029.
\newblock \doi{10.1103/PhysRevD.95.106002}.

\bibitem[Freidel and Livine(2019)]{freidelBubbleNetworksFramed2019}
Laurent Freidel and Etera~R. Livine.
\newblock Bubble {{Networks}}: {{Framed Discrete Geometry}} for {{Quantum Gravity}}.
\newblock \emph{General Relativity and Gravitation}, 51\penalty0 (1):\penalty0 9, January 2019.
\newblock ISSN 0001-7701, 1572-9532.
\newblock \doi{10.1007/s10714-018-2493-y}.

\bibitem[Freidel et~al.(2020)Freidel, Geiller, and Pranzetti]{freidelEdgeModesGravity2020}
Laurent Freidel, Marc Geiller, and Daniele Pranzetti.
\newblock Edge modes of gravity. {{Part II}}. {{Corner}} metric and {{Lorentz}} charges.
\newblock \emph{Journal of High Energy Physics}, 2020:\penalty0 27, November 2020.
\newblock ISSN 1029-8479.
\newblock \doi{10.1007/JHEP11(2020)027}.

\bibitem[Freidel et~al.(2021{\natexlab{a}})Freidel, Geiller, and Pranzetti]{freidelEdgeModesGravity2021}
Laurent Freidel, Marc Geiller, and Daniele Pranzetti.
\newblock Edge modes of gravity. {{Part III}}. {{Corner}} simplicity constraints.
\newblock \emph{Journal of High Energy Physics}, 2021\penalty0 (2007.12635):\penalty0 1--64, January 2021{\natexlab{a}}.
\newblock \doi{10.1007/JHEP01(2021)100}.

\bibitem[Engle et~al.(2010)Engle, Noui, Perez, and Pranzetti]{engleBlackHoleEntropy2010}
Jonathan Engle, Karim Noui, Alejandro Perez, and Daniele Pranzetti.
\newblock Black hole entropy from an {{SU}}(2)-invariant formulation of {{Type I}} isolated horizons.
\newblock \emph{Physical Review D}, 82\penalty0 (4):\penalty0 044050, August 2010.
\newblock ISSN 1550-7998, 1550-2368.
\newblock \doi{10.1103/PhysRevD.82.044050}.

\bibitem[Bodendorfer et~al.(2014)Bodendorfer, Thiemann, and Thurn]{Bodendorfer:2013jba}
Norbert Bodendorfer, Thomas Thiemann, and Andreas Thurn.
\newblock New {{Variables}} for {{Classical}} and {{Quantum Gravity}} in all {{Dimensions V}}. {{Isolated Horizon Boundary Degrees}} of {{Freedom}}.
\newblock \emph{Class. Quant. Grav.}, 31\penalty0 (5):\penalty0 055002, February 2014.
\newblock ISSN 0264-9381, 1361-6382.
\newblock \doi{10.1088/0264-9381/31/5/055002}.

\bibitem[Dupuis et~al.(2020)Dupuis, Freidel, Girelli, Osumanu, and Rennert]{Dupuis:2020ndx}
Ma{\"i}t{\'e} Dupuis, Laurent Freidel, Florian Girelli, Abdulmajid Osumanu, and Julian Rennert.
\newblock On the origin of the quantum group symmetry in 3d quantum gravity.
\newblock June 2020.
\newblock \doi{10.48550/arXiv.2006.10105}.

\bibitem[Perez(2013)]{perezSpinFoamApproach2013}
Alejandro Perez.
\newblock The {{Spin Foam Approach}} to {{Quantum Gravity}}.
\newblock \emph{Living Reviews in Relativity}, 16\penalty0 (1):\penalty0 3, December 2013.
\newblock ISSN 2367-3613, 1433-8351.
\newblock \doi{10.12942/lrr-2013-3}.

\bibitem[Speranza(2022)]{Speranza:2022lxr}
Antony~J. Speranza.
\newblock Ambiguity resolution for integrable gravitational charges.
\newblock \emph{JHEP}, 07\penalty0 (7):\penalty0 029, July 2022.
\newblock ISSN 1029-8479.
\newblock \doi{10.1007/JHEP07(2022)029}.

\bibitem[Barnich and Compere(2008)]{barnichSurfaceChargeAlgebra2008}
Glenn Barnich and Geoffrey Compere.
\newblock Surface charge algebra in gauge theories and thermodynamic integrability.
\newblock \emph{Journal of Mathematical Physics}, 49\penalty0 (4):\penalty0 042901, April 2008.
\newblock ISSN 0022-2488, 1089-7658.
\newblock \doi{10.1063/1.2889721}.

\bibitem[Freidel et~al.(2021{\natexlab{b}})Freidel, Oliveri, Pranzetti, and Speziale]{Freidel:2021cjp}
Laurent Freidel, Roberto Oliveri, Daniele Pranzetti, and Simone Speziale.
\newblock Extended corner symmetry, charge bracket and {{Einstein}}'s equations.
\newblock \emph{JHEP}, 09\penalty0 (9):\penalty0 083, September 2021{\natexlab{b}}.
\newblock ISSN 1029-8479.
\newblock \doi{10.1007/JHEP09(2021)083}.

\bibitem[Freidel et~al.(2021{\natexlab{c}})Freidel, Oliveri, Pranzetti, and Speziale]{freidelWeylBMSGroup2021}
Laurent Freidel, Roberto Oliveri, Daniele Pranzetti, and Simone Speziale.
\newblock The {{Weyl BMS}} group and {{Einstein}}'s equations.
\newblock \emph{Journal of High Energy Physics}, 2021\penalty0 (2104.05793):\penalty0 170, July 2021{\natexlab{c}}.
\newblock ISSN 1029-8479.
\newblock \doi{10.1007/JHEP07(2021)170}.

\bibitem[Cattaneo et~al.(2020)Cattaneo, Mnev, and Reshetikhin]{cattaneoCellularTopologicalField2020}
Alberto~S. Cattaneo, Pavel Mnev, and Nicolai Reshetikhin.
\newblock A {{Cellular Topological Field Theory}}.
\newblock \emph{Communications in Mathematical Physics}, 374\penalty0 (2):\penalty0 1229--1320, March 2020.
\newblock ISSN 1432-0916.
\newblock \doi{10.1007/s00220-020-03687-3}.

\bibitem[Baez(1998)]{baezSpinNetworksNonperturbative1998}
John Baez.
\newblock Spin {{Networks}} in {{Nonperturbative Quantum Gravity}}.
\newblock \emph{J. High Energy Phys.}, June 1998.
\newblock ISSN 9780821803806.
\newblock \doi{10.1090/psapm/051/1372769}.

\bibitem[Freidel et~al.(2023)Freidel, Geiller, and Wieland]{freidelCornerSymmetryQuantum2023}
Laurent Freidel, Marc Geiller, and Wolfgang Wieland.
\newblock Corner symmetry and quantum geometry, February 2023.

\bibitem[Mercuri(2006)]{mercuriFermionsAshtekarBarberoConnection2006}
Simone Mercuri.
\newblock Fermions in the {{Ashtekar-Barbero}} connection formalism for arbitrary values of the {{Immirzi}} parameter.
\newblock \emph{Physical Review D}, 73\penalty0 (8):\penalty0 084016, April 2006.
\newblock ISSN 1550-7998, 1550-2368.
\newblock \doi{10.1103/PhysRevD.73.084016}.

\bibitem[Geiller et~al.(2021)Geiller, Goeller, and Merino]{geillerMostGeneralTheory2021}
Marc Geiller, Christophe Goeller, and Nelson Merino.
\newblock Most general theory of 3d gravity: Covariant phase space, dual diffeomorphisms, and more.
\newblock \emph{Journal of High Energy Physics}, 2021\penalty0 (2):\penalty0 120, February 2021.
\newblock ISSN 1029-8479.
\newblock \doi{10.1007/JHEP02(2021)120}.

\bibitem[Ashtekar et~al.(1998)Ashtekar, Corichi, and Zapata]{ashtekarQuantumTheoryGeometry1998}
Abhay Ashtekar, Alejandro Corichi, and Jose~A. Zapata.
\newblock Quantum {{Theory}} of {{Geometry III}}: {{Non-commutativity}} of {{Riemannian Structures}}.
\newblock \emph{Classical and Quantum Gravity}, 15\penalty0 (10):\penalty0 2955--2972, October 1998.
\newblock ISSN 0264-9381, 1361-6382.
\newblock \doi{10.1088/0264-9381/15/10/006}.

\bibitem[{Figueroa-O'Farrill} and Stanciu(1994)]{figueroa-ofarrillNonsemisimpleSugawaraConstructions1994}
J.~M. {Figueroa-O'Farrill} and S.~Stanciu.
\newblock Nonsemisimple {{Sugawara Constructions}}.
\newblock \emph{Physics Letters B}, 327\penalty0 (1-2):\penalty0 40--46, May 1994.
\newblock ISSN 03702693.
\newblock \doi{10.1016/0370-2693(94)91525-3}.

\bibitem[Salgado et~al.(2014)Salgado, Szabo, and Valdivia]{salgadoTopologicalGravityTransgression2014}
Patricio Salgado, Richard~J. Szabo, and Omar Valdivia.
\newblock Topological gravity and transgression holography.
\newblock \emph{Physical Review D}, 89\penalty0 (8):\penalty0 084077, April 2014.
\newblock ISSN 1550-7998, 1550-2368.
\newblock \doi{10.1103/PhysRevD.89.084077}.

\bibitem[Bacry et~al.(1970)Bacry, Combe, and Richard]{bacryGrouptheoreticalAnalysisElementary1970}
H.~Bacry, {\relax Ph}.~Combe, and J.~L. Richard.
\newblock Group-theoretical analysis of elementary particles in an external electromagnetic field.
\newblock \emph{Il Nuovo Cimento A (1965-1970)}, 67\penalty0 (2):\penalty0 267--299, May 1970.
\newblock ISSN 1826-9869.
\newblock \doi{10.1007/BF02725178}.

\bibitem[Schrader(1972)]{schraderMaxwellGroupQuantum1972}
Robert Schrader.
\newblock The {{Maxwell Group}} and the {{Quantum Theory}} of {{Particles}} in {{Classical Homogeneous Electromagnetic Fields}}.
\newblock \emph{Fortschritte der Physik}, 20\penalty0 (12):\penalty0 701--734, 1972.
\newblock ISSN 1521-3978.
\newblock \doi{10.1002/prop.19720201202}.

\bibitem[de~Azcarraga et~al.(2011)de~Azcarraga, Kamimura, and Lukierski]{azcarragaGeneralizedCosmologicalTerm2011}
Jose~A. de~Azcarraga, Kiyoshi Kamimura, and Jerzy Lukierski.
\newblock Generalized cosmological term from {{Maxwell}} symmetries.
\newblock \emph{Physical Review D}, 83\penalty0 (12):\penalty0 124036, June 2011.
\newblock ISSN 1550-7998, 1550-2368.
\newblock \doi{10.1103/PhysRevD.83.124036}.

\end{thebibliography}

\end{document}